# Design and Simulation of RF Modulated Thermionic Electron Gun

Victor Mytrochenko, Valentin Zhyglo and Volodymyr Kushnir

NSC KIPT, 1 Academichna str, Kharkiv, Ukraine

**Abstract**

We present design and simulation results on the electron source for 2856 MHz technological electron linac. This device is a combination of an RF gun and a DC diode gun. The designed gun consists of a thermionic cathode, which is mounted on the wall of the $TM_{010}$ cavity that is under negative high voltage potential relatively to an anode. The use of such an electron source makes it possible to exclude the high-voltage gun modulator and resonator for beam prebunching from the accelerator circuit. Duration and position of an electron beam pulse within a RF power supply pulse of the accelerating section can be adjusted with tailoring of RF pulse feeding the cavity. The paper describes the design and electrodynamics characteristics of the gun cavity and a filter, which allows to transfer microwave power to that cavity. The particles dynamics has been studied by simulation. It is shown that the beam parameters at the gun exit satisfy the stated requirements. At an average particle energy of 95 keV, the beam pulse current is 0.44 A, the emittance does not exceed 4.8 mm·mrad, and 70% of all electrons are concentrated in the 72° phase range.

1. INTRODUCTION

One of the elements that determine the output beam quality in electron linacs is the electron source - the electron gun. If the conventional diode electron gun is used, the formation of electron bunches, occurs in a bunching system and in first accelerating section. Therefore, the idea of obtaining a modulated particle flux at an electron gun output with modulation frequency equal to the operating linac frequency $f_0$, or its subharmonic, is very attractive.

Obviously, the requirements for an electron source are determined by the accelerator purpose and the requirements for the beam parameters at its output. In particular, in accelerator based free electron lasers, radiofrequency (RF) guns with cathodes of various types were designed. Thus, in the most widely used photocathode RF guns, precision beams with extremely high brightness are obtained [1, 2]. Although the thermionic RF guns are much simpler than the photocathode guns; in particular, they do not require the use of a laser system, they are suitable to obtain even femtosecond bunches using non-isochronous compression device [3].

Currently, intensive research is being carried out on electron sources that would combine the advantages of RF guns and DC guns with different cathodes. For example, a DC gun with photoemission cathode has been designed, built, and commissioned at Cornell University [4]. The high-voltage (350 kV) DC diode gun contains the GaAs photocathode, irradiated with the 1 ps duration laser pulses at pulse repetition rate of 1.3 GHz. This makes it possible to obtain at the injector output bunch charge of 77 pC with a normalized emittance of 0.53 μm. Paper [5] describes a combined gun with thermionic $CeB_6$ cathode. The cathode is installed in a complex form stripline (stripline loop). Two high voltage 1 ns pulses propagating in strip-loop towards each other created electric field on the cathode. Emitted electrons pass through the aperture of the line outer conductor and accelerated in the RF resonator. Such a gun, according to simulations, makes it possible to form beams with an emittance, close to the thermal level of 0.3 mm·mrad.

The various possibilities to obtain both pulsed and RF modulated beam near the cathode can be realized by using a triode electron gun. In the most used accelerator schemes the nanosecond pulses are applied to the grid. The pulse repetition rate is equal to one of subharmonics of $f_0$. Then, pulsed beam generated in the gun is compressed by subharmonic bunchers. The advantage of such schemes for the formation of bunches is the bunches production with a large charge. Such a bunching scheme is often used in synchrotron injector accelerators. An example is the linear accelerator-injector for ALBA synchrotron [6] with a triode 90 kV electron gun. In one of operating modes of this accelerator, pulses with duration of 1 ns are applied to the gun grid. Pulse duration of 1 ns corresponds to of 180° at the sixth subharmonic of the accelerator's operating frequency. Further, 0.25 nK bunch is compressed by bunchers to duration of 17 ps. A similar scheme of bunch formation is used in compact accelerators for various purposes including FEL [7, 8, 9].

In a triode gun, the beam RF modulation can also be carried out. This approach was first applied in the accelerator FELIX [10]. A microwave signal with frequency of 1 GHz (3rd subharmonic of the linac operating frequency) is applied to the grid of the 100 kV gun. By choosing ratio between the constant bias value and the RF field amplitude on the grid, one can control the bunch length at the output of the gun. It is possible to provide modulation at higher linac operating frequency using CPI Eimac cathode-grid assemblies. Such a gun, for example, was developed for the 2856 MHz linac at Pohang Accelerator Laboratory [11, 12]. At the same time, the using for near-cathode modulation of standard cathode assemblies at high frequency have some problems caused by both the grid configuration and the cathode-grid large capacity. The creation of a special near-cathode resonator with an appropriate microwave power supply system is one of the solutions to these problems. The article presents the results of calculation and numerical simulation of an electron gun that contains these elements.

## 2. COMBINED RF/DC GUN DESIGN

The developed gun is intended for use in the upgraded S band technological electron linac LU-10 [13]. It is an alternative version of the electron source [14], which has already been developed and manufactured for this accelerator. The injection system of the linac consists of a diode thermionic gun with a voltage of 80 kV and $TM_{010}$ prebuncher. A beam with a current of 0.4 A enters the accelerating section. The suggested gun (Figure 1) consists of a thermionic cathode, which is installed on the wall of the $TM_{010}$ cavity and anode, which is located at some distance from the cavity. The electron bunches, which are produced in the RF cavity, are rapidly accelerated in the space cavity-anode and then injected into the accelerating section. Thus, this device is a combination of RF gun and DC diode gun. The RF cavity is under high voltage potential relative to ground, so a filter that protects the microwave power source and supply lines from high voltage is required.

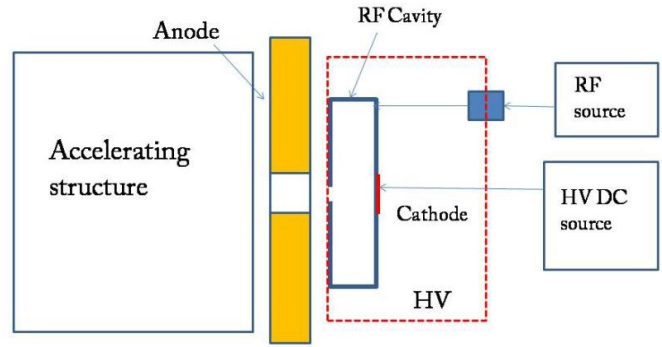

Figure 1: Layout of the Gun

The cavity tuned to frequency of 2856 MHz. The cavity diameter and the accelerating gap are 79.08 mm and 2mm, respectively. The hole diameter for the beam exit is 2 mm. The microwave power is supplied to the resonator by two diametrically located waveguides, which makes it possible to reduce the field asymmetry in the cathode region. According to the simulation, the loaded quality factor of the cavity is 1540, the shunt impedance is 2.68 M$\Omega$/m. With a high-frequency power of 11.9 kW supplied to the cavity, the average field on the axis is 1.7 MV/m. An anode is located at 10 mm from the cavity. The LaB$_6$ cathode is installed in the traditional way for thermionic RF guns using a $\lambda/2$ filter. To obtain the required output beam current, the cathode diameter was chosen to be 3 mm. Simulation shows that at an emission density of 18 A/cm$^2$, the cathode current in the operating mode is 1.3 A, and the current at the gun output is 0.44 A. The RF and DC fields are shown in Figure 2.

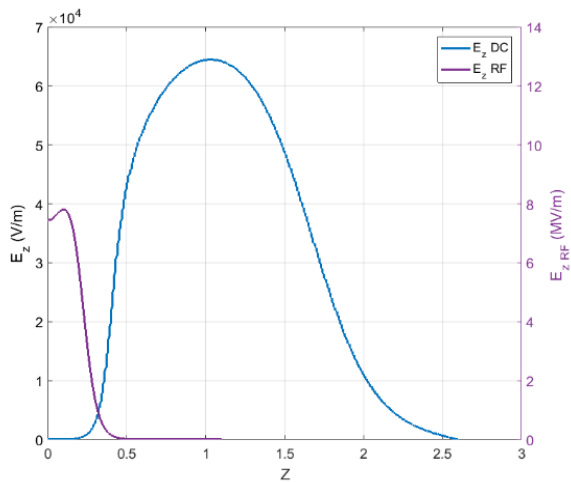

Figure 2: The distribution of RF and DC field on the gun axis

It can be seen that there is only a high-frequency field at the cathode. This makes it possible to abandon the high-voltage gun modulator since the current pulse duration at the gun output is determined by the RF pulse duration. This allows to adjust easily the duration of the pulse current using PIN diode RF switch, which is connected in circuit of the gun RF power supply.

The gun is powered both by a highly stable DC voltage source of 80 kV and by RF power source. A simplified diagram of the gun high-frequency power supply is shown in Figure 3. Using a waveguide adjustable directional coupler, part of the klystron power is fed through a phase shifter, the PIN diode RF switch and the filter to coaxial hybrid ring coupler and then through coaxial-to-wave guide adapters to 73x10 mm waveguides, which provide RF power to the cavity.

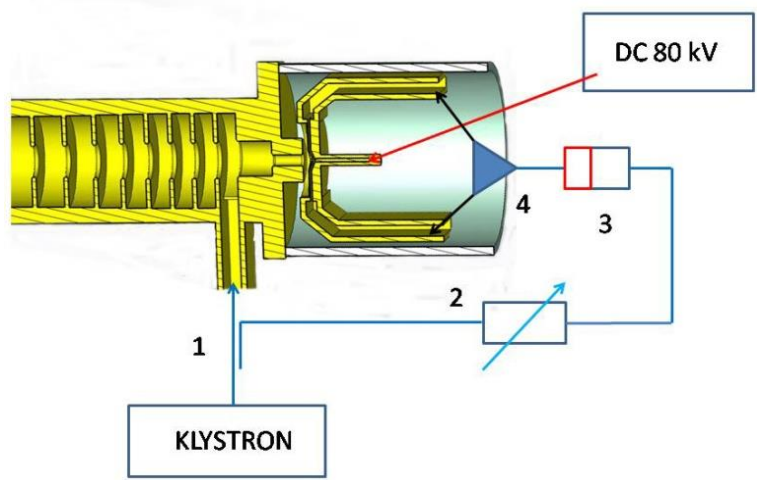

Figure 3: Schematic representation of the gun. 1 – waveguide directional coupler, 2 - variable phase shifter and the PIN diode RF switch, 3 – filter, 4- coaxial hybrid ring coupler

### 3. FILTER

Let us consider in more detail the developed filter, which should both ensure the transmission of RF power with minimal losses and perform an insulator function withstanding the voltage of 80 kV. The device (see Figure 4) consists of input rectangular waveguide 72x34 mm, the rectangular horn, the dielectric rod and the coaxial adapter. In such a system, several mode transformations take place. Thus, in the input device, the $TE_{10}$ wave in a rectangular waveguide is transformed into a hybrid $ME_{11}$ mode in the dielectric rod. At the output device, the $ME_{11}$ mode is transformed into a $TE_{11}$ cylindrical mode, which is transformed into a TEM mode in a coaxial. The material of the dielectric rod must have low dielectric losses (for example, Teflon, polystyrene, quartz, or RF ceramics), good insulating properties and sufficiently high radiation resistance. We chose quartz ($\varepsilon$=3,75, tg$\delta$=1·10$^{-4}$), which has all these properties for simulation and filter design. The diameter of the rod regular part (48 mm) was determined to maintain the single-mode regime in a quartz dielectric waveguide [15].

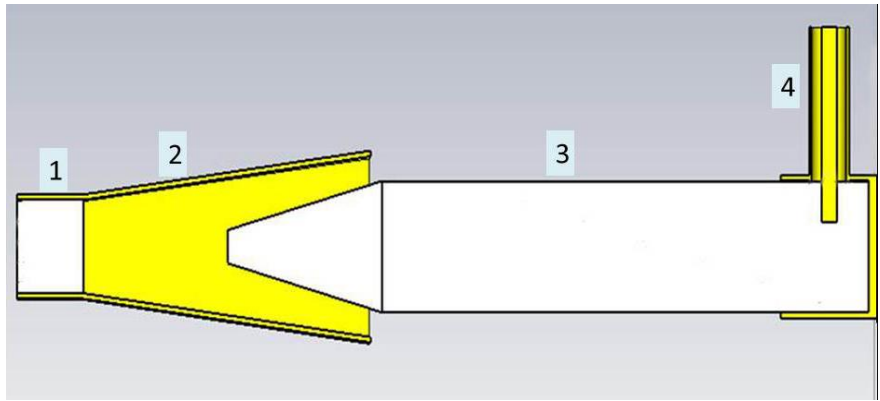

Figure 4: Schematic of the filter. 1 - rectangular $TE_{10}$ waveguide, 2 - rectangular horn, 3 - section of dielectric waveguide, 4 - transitions from a cylindrical waveguide to coaxial line

The rod length of 220 mm ensures reliable insulation. The main sources of microwave power losses in such a system are reflection losses, losses in metal walls and dielectric, and radiation losses from the surface of the dielectric rod. Analysis shows that most of the losses are due to the presence of numerous reflections from inhomogeneities within the junction. The device was adjusted by sequentially changing the dimensions of its elements. As a result, at frequency of 2856 MHz, the values of $S_{11}$=-19 dB and $S_{12}$=-1.04 dB were obtained, i.e., the standing wave ratio at the input is 1.25 and the power loss is about 20%.

## 4. BEAM DYNAMICS SIMULATION

The particle dynamics in the gun was simulated using the Parmela program [16]. In modeling, it was assumed that the initial bunch length is 360°, which corresponds to the continuous emission of particles from the cathode surface, and the initial energy of electrons at the cathode is 0.15 eV. Electrons enter the cavity only during the half-cycle of microwave oscillations. The simulation was carried out at various phases of the field in the cavity and at various reference particle positions. In all cases the main beam parameters at the gun exit changed insignificantly and difference does not exceed of the simulation error. Therefore, we present below the results obtained with the cavity phase of 180°. In this case, the second half of the injected bunch is accelerated, and the parameter $Z_0$ is chosen so that the reference particle enters the resonator at a phase of 180°. The beam dynamic simulation showed that at cathode current of 1.3 A, the gun output current is 0.444 A, i.e., 66% of the injected particles come back. The energy and phase characteristics of the beam at the gun exit are shown in Figure 5. The average and maximum electron energies at the gun output are 95 keV and 103 keV, respectively. In this case, for 70% of the particles, the relative energy spread does not exceed 11%. In the longitudinal phase space, the beam is a compact bunch and 70% of the particles are concentrated within 72°. This provides a good particle capture rate in the acceleration process in the accelerating section. The spatial characteristics of the beam at the gun exit are shown in Figure 6.

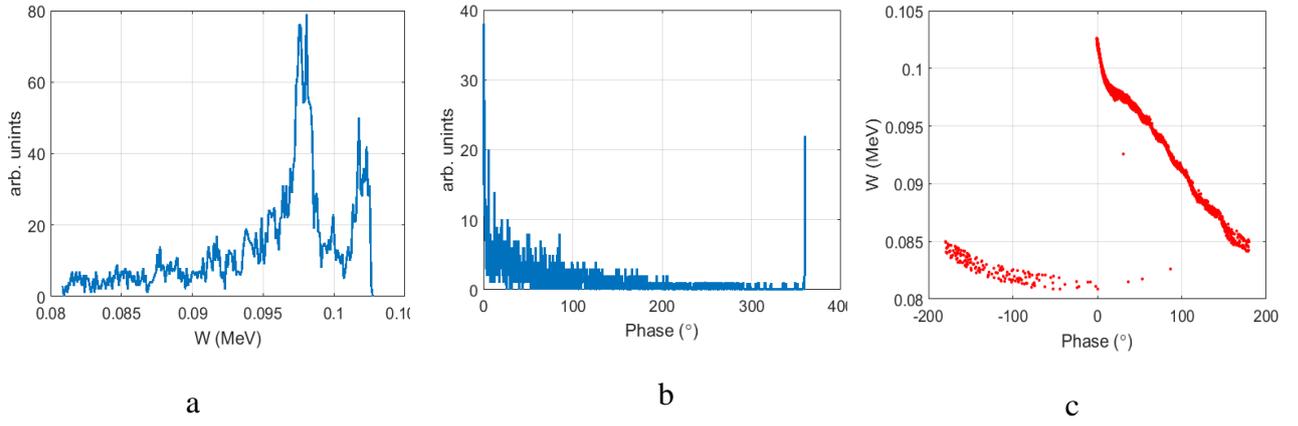

Figure 5: Energy spectrum (a), phase spectrum (b) and phase-energy distribution of electrons at the gun output (c)

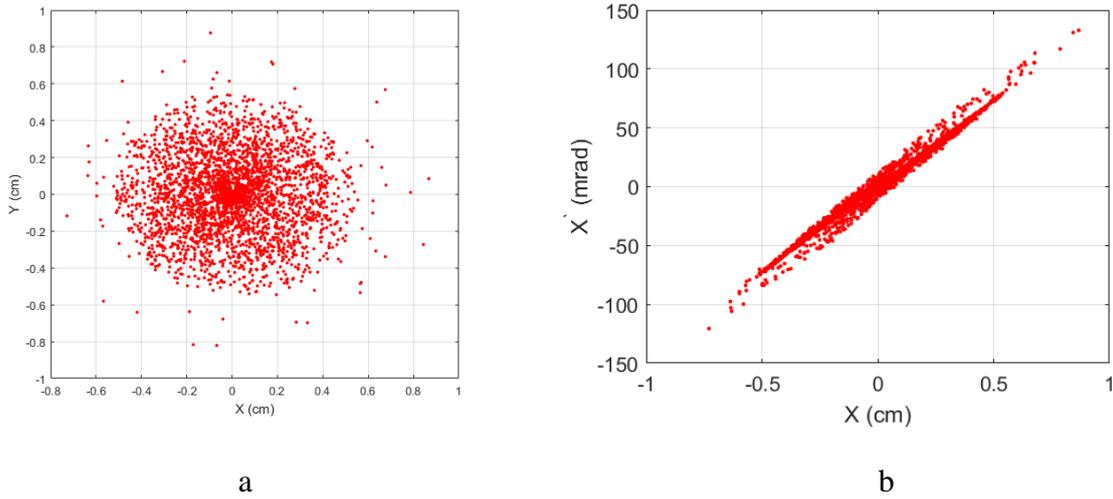

Figure 6: Distribution of electrons on the transverse XY plane (a) and on the phase XX´ plane (b)

It can be seen that the distribution of beam particles in the transverse plane is close to symmetrical. The rms beam diameters are $D_x$=4.67 mm and $D_y$=4.63 mm. The diameters for 95% of particles in the X and Y directions are 8.81 mm and 8.77 mm, respectively. The value of the rms normalized beam emittance in different planes is also close: $\varepsilon_x$=4.79 mm·mrad and the $\varepsilon_y$=4.70 mm·mrad. The slight asymmetry in the spatial distribution of the particle density observed in the simulation is apparently caused by RF field asymmetry in the cavity. The output beam is divergent (see Figure 6), but the divergence is such that it allows the beam to be injected into the accelerating section without loss of particles. The time for current to reach its steady state depends on cavity filling time and does not exceed 0.2 μs. This value is acceptable at the required current pulse width of 10 μs.

As is known, one of the problems in the existing thermionic RF guns is the back bombardment of the cathode. This effect limits the possibility of both increasing the pulse duration and increasing the pulse repetition rate. In our case, power of the reversed electrons is significantly less than the cathode heating power even at a pulse repetition rate of 600 Hz. Therefore, the back bombarding electrons do not significantly change the cathode temperature. Compared to the previously developed LU-10 injection system (the diode gun and the prebunching resonator), the cathode microwave modulation gun has significantly better beam parameters. In particular, the bunch duration is 2.5 times smaller.

**SUMMARY**

The results of calculations and numerical simulation showed the possibility of creating the electron gun for S band linac with near cathode RF modulation. The beam at the gun output is a sequence of bunches. The bunch repetition frequency is equal to the linac operating frequency. The beam parameters at the gun exit fully satisfy the stated requirements. Its application makes it possible to exclude the high-voltage electron source modulator and the resonator prebuncher from the accelerator circuit and improve the beam parameters at the linac output. The developed gun is designed for a specific technological accelerator, but the same principle - the principle of near-cathode microwave beam modulation by RF cavity can also be used in accelerators designed for scientific research.